\documentclass{webofc}
\usepackage[varg]{txfonts}   % Web of Conferences font
\usepackage{hyperref,float,comment}
\usepackage{url}
\hypersetup{colorlinks=true,citecolor=blue,urlcolor=blue,linkcolor=blue}
\def \bes {\begin{subequations} }
\def \ees {\end{subequations}}
\def \be{\begin{eqnarray}}
\def \ee{\end{eqnarray}}
\begin{document}
\title{Factorial cumulants of proton multiplicity near a critical point using maximum entropy freeze-out prescription}
\author{\firstname{Jamie} \lastname{Karthein}\inst{1}\fnsep%\thanks{\email{Mail address for first
   % author}} 
    \and
        \firstname{Maneesha} \lastname{Pradeep}\inst{2}\fnsep\thanks{\email{mpradeep@umd.edu}}\and\firstname{Krishna} \lastname{Rajagopal}\inst{1}\fnsep%\thanks{\email{Mail address for second
             %author if necessary}} 
             \and
        \firstname{Mikhail} \lastname{Stephanov}\inst{3}\fnsep%\thanks{\email{Mail address for last
             %author if necessary}}
             \and
        \firstname{Yi} \lastname{Yin}\inst{4}\fnsep%\thanks{\email{Mail address for last
             %author if necessary}}
        % etc.
}

\institute{Massachusetts Institute of Technology, Cambridge, MA 02139, USA 
\and
            University of Maryland, College Park, MD 20742 USA
\and
           University of Illinois, Chicago, Illinois 60607, USA
\and   
            The Chinese University of Hong Kong,
Shenzhen, Guangdong 518172, China
          }

%\author{\firstname{Maneesha } \lastname{Pradeep}\inst{1,3}\fnsep\thanks{\email{mpradeep@umd.edu}}%\and
 %       \firstname{Isabelle} \lastname{Houlbert}\inst{2}\fnsep\thanks{\email{Mail address for second
 %            author if necessary}
 %} 
 %\and
    %    \firstname{Agnès} %\lastname{Henri}\inst{3}\fnsep\thanks{\email{Mail address for last
     %        author if necessary}}
        % etc.
%}
%}
%\institute{University of Maryland, College Park}
%\and
           
%\and
%           Last address
%\and           
     %       Address

\abstract{%A key question about the QCD phase diagram is whether there is a critical point somewhere on the boundary between the hadron gas and quark-gluon plasma phases, and if so where. Heavy-ion collisions offer a unique opportunity to search for signatures of such a critical point by analyzing event-by-event fluctuations in particle multiplicities. To draw meaningful conclusions from experimental data, a theoretical framework is needed to link QCD thermodynamics with the particle spectra and correlations observed in detectors. The unknown Equation of State (EoS) of QCD near a critical point can be related to the universal Gibbs free energy of the 3D Ising model using four non-universal mapping parameters whose values are determined by the microscopic details of QCD. We use maximum entropy freeze-out of fluctuations to make estimates for the factorial cumulants of proton multiplicities, assuming thermal equilibrium.
%A key question in the QCD phase diagram is whether a critical point exists on the boundary between the hadron gas and quark–gluon plasma. Heavy-ion collisions probe this through event-by-event multiplicity fluctuations, which require a theoretical framework connecting QCD thermodynamics with observed spectra. Mapping the QCD Equation of State near criticality to the universal 3D Ising free energy, we use maximum-entropy freeze-out to estimate factorial cumulants of proton multiplicities in thermal equilibrium.
We present the first application of the maximum-entropy freeze-out prescription to calculate factorial cumulants of proton multiplicities near the conjectured QCD critical point in thermal equilibrium. We map the Gibbs free energy of the 3D Ising model to a parameterized class of possible EoS near QCD critical point.  This equilibrium baseline highlights how factorial cumulants isolate critical fluctuations by subtracting trivial self-correlations, setting the stage for future out-of-equilibrium analyses. We identify the key nonuniversal aspects of the mapping to the Ising model that strongly control the characteristic properties, such as magnitude and location of the peaks of the factorial cumulants along the freeze-out curve.
}
\maketitle
\section{Introduction}
\label{intro}
\hspace{1cm}A central open question in QCD is whether a critical point lies on the boundary between the hadron resonance gas and the quark–gluon plasma. Event-by-event particle multiplicity fluctuations are among the most sensitive probes of such a critical point, with factorial cumulants being particularly useful since they subtract trivial self-correlations of a quantum gas. The STAR collaboration has published precision measurements of the factorial cumulants of proton multiplicity as a function of collision energy \cite{STAR:2025zdq} . The data for the normalized second and third factorial cumulants of proton multiplicity show statistically significant deviations from the computed non-critical baselines \cite{Vovchenko:2025jgy}. To interpret such data and constrain the QCD equation of state near criticality, a framework is needed that connects QCD thermodynamics to the particle spectra measured after freeze-out.

\hspace{1cm}In Ref. \cite{Karthein:2025hvl}, we take a first step in this direction. Using a map from the universal 3D Ising model to a family of candidate QCD equations of state\cite{Parotto:2018pwx}, we apply the maximum-entropy freeze-out prescription \cite{Pradeep:2022eil} to compute factorial cumulants of proton multiplicities in thermal equilibrium. This provides a controlled equilibrium baseline for future studies that incorporate the out-of-equilibrium dynamics expected near the critical point. Although finite-time effects are expected to become increasingly important near the critical point—making an equilibrium based calculation incomplete for heavy-ion collisions—it serves as a necessary prerequisite to more sophisticated out-of-equilibrium treatments.

\hspace{1cm}%The qualitative behavior of baryon-number susceptibilities  along a freeze-out curve passing near a critical point \cite{Stephanov:2024xkn} has been long anticipated and has recently been revisited in light of new experimental results. 
Since protons carry both baryon number and energy, there factorial cumulants of proton multiplicity depend on the cumulants of baryon density as well those of energy density. The maximum-entropy prescription for freeze-out\cite{Pradeep:2022eil} maximizes the relative entropy of the hadron resonance gas (i.e. the entropy of fluctuations relative to those of an uncorrelated HRG) while ensuring that local hydrodynamic densities are matched across the kinetic and hydrodynamic descriptions at freeze-out on an event-by-event basis. %In this proceedings, we briefly summarize the results from Ref.~\cite{Karthein:2025hvl} where we apply this method under the simplifying assumption of thermal equilibrium at freeze-out. Although finite-time effects are expected to become increasingly important near the critical point—making an equilibrium based calculation incomplete for heavy-ion collisions—it serves as a necessary prerequisite to more sophisticated out-of-equilibrium treatments. We further explore the dependence of the proton factorial cumulants on the non-universal parameters of the mapped EoSs and on the distance of the freeze-out location from the critical point on the QCD phase diagram. 

\section{QCD Equation of State near the critical point from universality}

\begin{figure}[h]
\centering
\sidecaption
\includegraphics[width=4cm,clip]{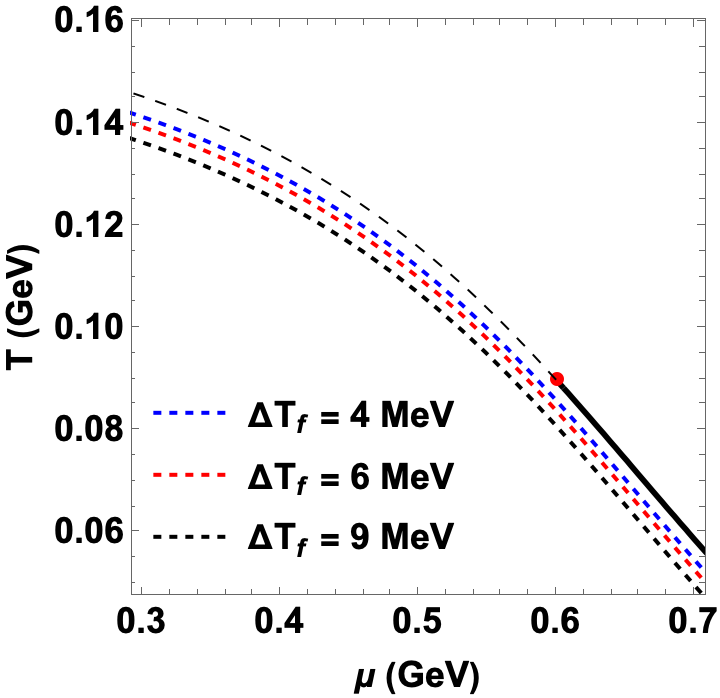}
\caption{ Three freezeout curves displaced downward relative to the crossover curve $\Delta T'=0$ (black) by $\Delta T_f=4$, 6 and 9 MeV (dashed blue, red and black curves, respectively). An important non-equilibrium effect is that the fluctuations at freeze-out remember conditions at an earlier higher temperature, so it is possible that choosing a small $\Delta T_f$ in our calculation is a reasonable representation for freeze-out at a lower temperature in a calculation that includes non-equilibrium effects. For this reason $\Delta T_f$  in our calculation should be considered a model parameter to be fit to data. }
\label{fig-0}       % Give a unique label
\end{figure}

\hspace{1cm}The EoS of QCD at non-vanishing densities, and the phase diagram of QCD are less understood from first principles due to the notorious sign problem related to lattice QCD calculations. If the conjectured critical point of QCD exists, it would belong to the 3D Ising universality class. The universality near the critical point implies that, in the vicinity of the QCD critical point, its thermodynamic behavior can be described by that of the 3D Ising model via a non-universal mapping (with six parameters) between the QCD and Ising scaling fields.%The universality implies that the EoS of QCD close to the critical point can be mapped to 3D Ising model via six non-universal (microscopic) parameters. %The linear mapping between the QCD variables , temperature $(T)$ and baryon chemical potential $(\mu)$ and the Ising variables, Ising reduced temperature $(r)$ and magnetic field $(h)$ is parametrized as below:
%\bes
%\label{Eq:linmap}
%\be \label{Eq:linmap1}
%h(\mu,T)&=&-\frac{\Delta T \cos \alpha_1 +\Delta \mu \sin \alpha_1}{w T_c \sin (\alpha_{1}-\alpha_2)}\\ \label{Eq:linmap2}
%r(\mu,T)&=&\frac{\Delta T \cos \alpha_2 +\Delta \mu \sin \alpha_2}{\rho w\, T_c \sin (\alpha_{1}-\alpha_2)}\ .
%\ee
%\ees
%where $\Delta T=T-T_c, \Delta \mu=\mu-\mu_c$ and $(T_c, \mu_c)$ are the values of temperature and baryon chemical potential at the critical point. $\alpha_1, \alpha_2$ are the angles that Ising axes $h=0$ and $r=0$ make with the $-\mu$ axis. $w$ and $\rho$ are scaling variables that determine how Ising $h$ and $r$ axes scale with $\Delta T$ and $\Delta \mu$. 
The six non-universal parameters correspond to the values of the critical chemical potential, $\mu_c$, the critical temperature, $T_c$, the angles that $r$ and $h$ axis make with the QCD $\mu$ axis on the $T-\mu$ phase diagram, denoted by $\alpha_1$ and $\alpha_2$ respectively and two scale factors $\rho$ and $w$ \cite{Parotto:2018pwx}. In this proceedings we make the following choices for two of the non-universal parameters: $\mu_c\,=\, 600\, \text{MeV}$ and $\alpha_2\,=\, 0^\circ$. The values of the critical temperature,$T_c$ and the slope of the first order line, $\alpha_1$ then follow from the pseudo-critical boundary obtained from lattice calculations\cite{Kahangirwe:2023ynz}. The critical point, the phase boundary and the freeze-out curves used for the analysis presented in this proceedings are shown in Fig.~(\ref{fig-0}).

The hydrodynamic correlations in equilibrium can be directly calculated from the Equation of State. Ref.\cite{Pradeep:2022eil} presents a way to calculate the non-trivial correlations in a quantum gas of hadrons from the hydrodynamic correlations using maximum entropy freeze-out. For different values of the mapping parameters $\rho$ and $w$, we study how the factorial cumulants of proton multiplicities vary along different freeze-out curves specified by
\be\label{eq:FreezeoutCurve}
T_f(\mu_B)=T_{\rm crossover}(\mu_B)-\Delta T_f
\ee
with $\Delta T_f=4$, 6 and 9 MeV, where $T_{\rm crossover}(\mu_B)$ is the Ising-$r$ axis where $h=0$. 
\begin{comment}
The Maximum Entropy freezeout prescription for the factorial cumulants of the proton multiplicities, normalized by the mean proton multiplicity, $\left<N\right>$ , denoted by $\hat{\Delta}\omega^k$, is given by:
\be\label{Eq:IRCOmegaSimple}
\hat{\Delta}\omega_{k}=\hat{\Delta}H_{a_1\dots a_k}P^{a_1}\dots P^{a_k}\,\left<N\right>^{-1}
\ee
where the $P^{a}$'s are given by 
\be
\label{Eq:Xa}
P^{a}_{\hA}\equiv \int_{\widetilde A}P^a
=\int_{\tilde{A}}\, (\bar{H}^{-1})^{ab}\,P_{b}^{B} \,\bar{G}_{BA} %= \bar{H}^{-1^{ab}}\bar{\<P_{bA}\>}\ .
\
\end{comment}

%For figure with sidecaption legend use syntax of figure~\ref{fig-3}

\section{Results on the phase diagram}

\begin{figure}[H]
\centering
\includegraphics[width=12cm,clip]{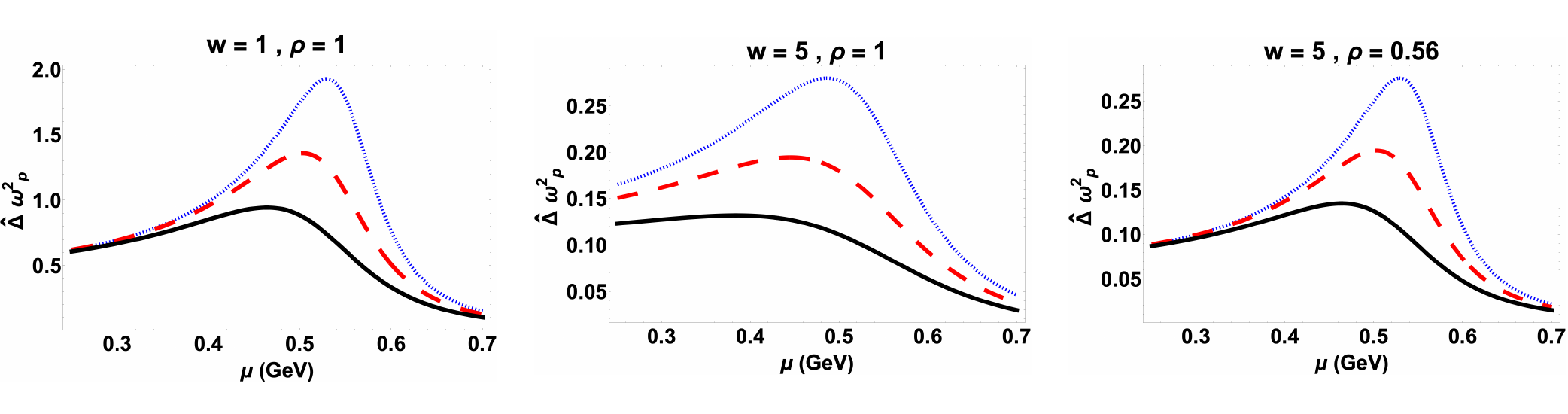}
\caption{The second factorial cumulant of the proton multiplicity distribution, $\hat{\Delta}\omega_{2p}$, along the three freezeout curves from Fig.~\ref{fig-0} characterized by Eq.~\eqref{eq:FreezeoutCurve} with $\Delta T_f=4$, 6 and 9 MeV (blue dashed, red dashed and black dotted respectively). The panels show $\hat\Delta\omega_{2p}$ 
    for various values of the nonuniversal mapping parameters $w$ and $\rho$ (specified on top of the figure), with $\mu_c=600 \, \text{MeV}$, $T_c=90$~MeV and  $\alpha_2=0^{\circ}$. The peak height decreases as $w$ increases, consistent with the scaling $w^{-6/5}$, while the location of the peak is controlled by the quantity $\bar{\rho}=\rho w^{2/5}$. }
\label{fig-1}       % Give a unique label
\end{figure}

\begin{figure}[h]
\centering
\includegraphics[width=12cm,clip]{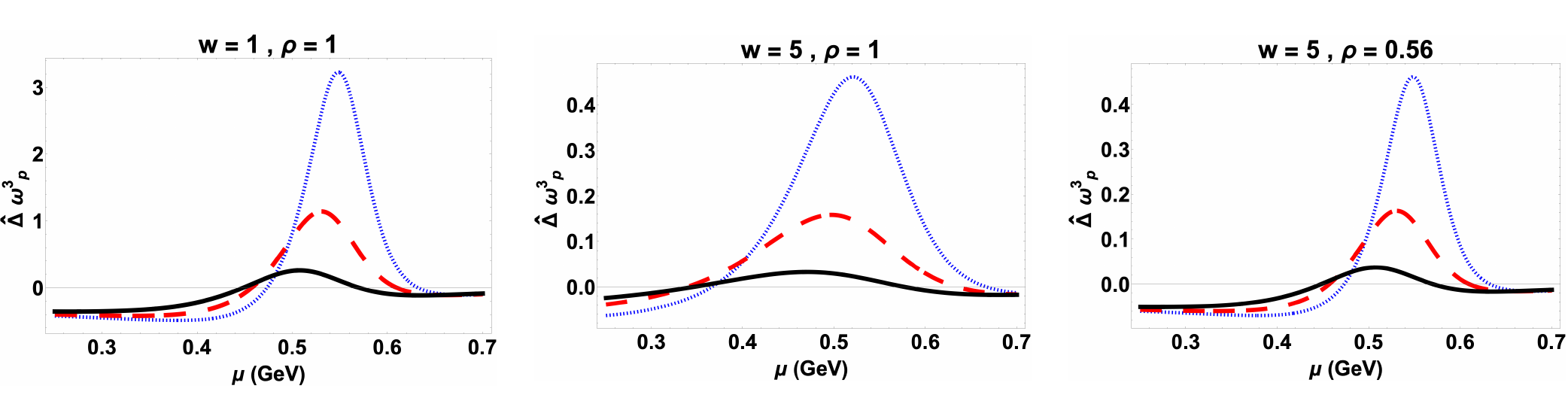}
\caption{The third factorial cumulant of the proton multiplicity distribution, $\hat{\Delta}\omega_{3p}$, along the three freezeout curves from Fig.~\ref{fig-0} characterized by Eq.~\eqref{eq:FreezeoutCurve} with $\Delta T_f=4$, 6 and 9 MeV (blue dashed, red dashed and black dotted respectively). }
\label{fig-3}       % Give a unique label
\end{figure}

For the presentation of the results here, we chose $(w,\rho)$ values $(1,1)$, $(5,1)$ and $(5,0.56)$ and $\Delta T_f$ values as specified in Fig.~(\ref{fig-0}).  
The key observation is that while the position of the peak of the factorial cumulants of proton multiplicity along the freeze-out curve is controlled by the combination $\bar{\rho}=\rho w^{1-\frac{1}{\beta\delta}}$ \cite{Karthein:2025hvl,Basar25}, the magnitude of the maximum value is more sensitive to $w$, and it goes as $w^{-1-1/\delta}$ , where $\beta$ and $\delta$ are the critical exponents for 3D Ising universality class. ($\beta\approx 0.326$ and $\delta\approx 4.8$.) This can be seen in Figs.~(\ref{fig-1},\ref{fig-3}), where the second and third factorial cumulant of the proton multiplicity distribution, along the three freeze-out curves have been plotted. The $\rho=0.56$
value for the rightmost panel in each of these plots have been chosen such that the $\bar{\rho}$ remains the same between the leftmost and rightmost plots while $w$ changes. The magnitude of the maxima of the $k^{\text{th}}$ factorial cumulant of proton multiplicity goes as $\Delta T_f^{1+1/\delta-k}$ \cite{Karthein:2025hvl}.

\begin{figure}[H]
\centering
\includegraphics[width=12cm,clip]{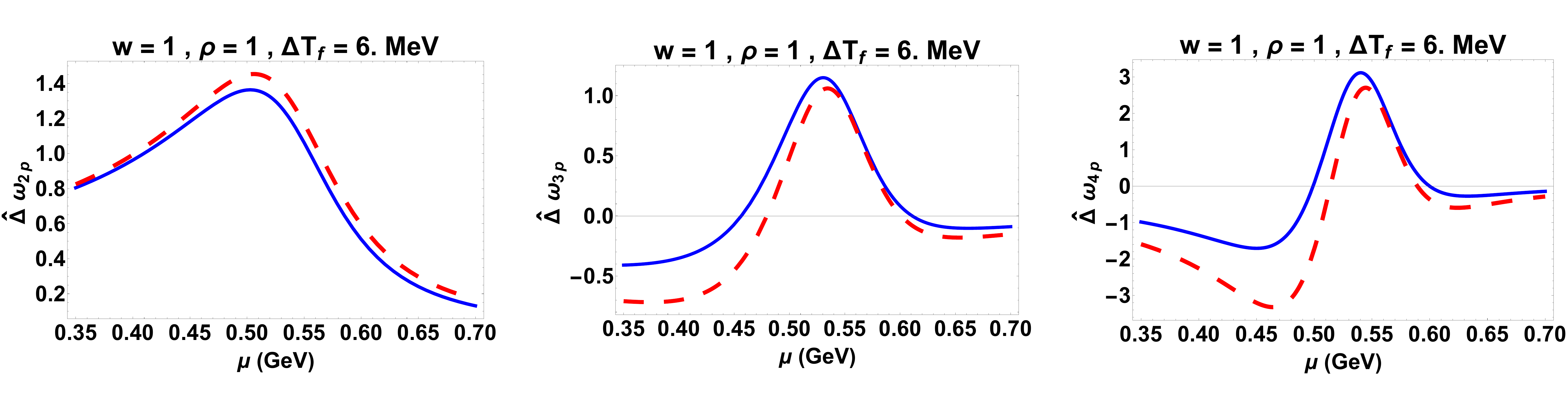}
\caption{The second, third and fourth factorial cumulants of the proton multiplicity distribution, along the freezeout curve of Eq.~(\ref{eq:FreezeoutCurve}) with $\Delta T_f =6$~MeV. The red dashed curves show $\hat\Delta\omega_{4p}$,  which includes contributions of direct and child protons, whereas the blue solid curves include 
    only the direct protons, as in Figs.~\ref{fig-1} and \ref{fig-3}.}
\label{fig-4}       % Give a unique label
\end{figure}

Figs.~(\ref{fig-1},\ref{fig-3}), include only the contribution of direct protons, i.e those protons which existed at the time of freeze-out. In experiments, both direct protons as well as those coming from the decay products of heavier resonances contribute to the observed proton multiplicity distribution. %Maximum entropy freeze-out allows to estimate the contribution of the child (decay) protons as well into the calculation of factorial cumulants. 
We find that the inclusion of the child (decay) protons do change the estimates quantitatively, but donot change the qualitative trends. A comparison between the calculation including only direct protons and the one including both direct and child protons is shown for the choice of $\rho=w=1$ and $\Delta T_f=6\, \text{MeV}$ in Fig.~(\ref{fig-4}).

\section{Looking forward}

Ref. \cite{Karthein:2025hvl}, whose main results are summarized in this proceedings, presents the first application of the maximum-entropy freeze-out method to estimate experimentally accessible fluctuation observables—specifically, the factorial cumulants of proton multiplicity—for parametrically specified trial QCD equations of state. While this work assumes thermal equilibrium, more realistic calculations incorporating out-of-equilibrium effects remain an important next step. Looking ahead, a Bayesian comparison between experimental measurements of proton factorial cumulants and our theoretical baselines will provide quantitative constraints on the non-universal mapping parameters and, crucially, on the location of the QCD critical point.

\section{Acknowledgements}
This work was supported by the U.S.~Department of Energy, Office of Science, Office of Nuclear Physics under grants DE-SC0011090, DE-FG02-93ER40762, and DE-FG02-01ER41195.  JMK is supported by an Ascending Postdoctoral Scholar Fellowship from the National Science Foundation under Award No. 2138063. YY acknowledges the support from NSFC under Grant No.12175282 and from CUHK-Shenzhen University Development Fund under the Grant No. UDF01003791.
%For two-column wide figures use syntax of figure~\ref{fig-2}
%\begin{figure*}
%\centering
%\vspace*{1cm}       % Give the correct figure height in cm
%\includegraphics[width=8cm,clip]{fig-2-sample}
%\label{fig-2}       % Give a unique %label
%\end{figure*}

%For figure with sidecaption legend use syntax of figure~\ref{fig-3}
%\begin{figure}
%\centering
%\sidecaption
%\includegraphics[width=5cm,clip]{fig-3-sample}
%\label{fig-3}       % Give a unique label
%\end{figure}

%
% BibTeX or Biber users please use (the style is already called in the class, ensure that the "woc.bst" style is in your local directory)
% \bibliography{your_bib_file} % Replace "your_bib_file" with the actual name of your .bib file

\end{document}